\documentclass[12pt]{article}
\usepackage{float} 
\usepackage{amsmath}
\usepackage{slashed}
\usepackage{amsfonts}   
\usepackage{amssymb}

\begin{document}
\date{}
\title{{\bf{\Large Holographic derivation of $ q $ SYK spectrum with Yang-Baxter shift}}}
\author{
 {\bf {\normalsize Dibakar Roychowdhury}$
$\thanks{E-mail:  dibakarphys@gmail.com, dibakarfph@iitr.ac.in}}\\
 {\normalsize  Department of Physics, Indian Institute of Technology Roorkee,}\\
  {\normalsize Roorkee 247667 Uttarakhand, India}
\\[0.3cm]
}

\maketitle
\begin{abstract}
In this Letter, based on the notion of gauge/gravity duality, we compute $ q $ SYK spectra in the presence of Yang-Baxter (YB) deformations. The gravitational counterpart of this duality turns out to be the YB deformed Almheiri-Polchinski (AP) model embodied with hyperbolic potential for the dilaton. Following Kaluza-Klein reduction on $ (AdS_2)_{\eta}\times (S^1/ Z_2) $, we compute the deformed Green's function that reproduces the original $ q $ SYK spectrum in the limit of vanishing deformations.
\end{abstract}
\section{Overview and motivation}
Recently, the Sachdev-Ye-Kitaev (SYK) model \cite{Sachdev:1992fk}-\cite{Sachdev:2015efa} has been proposed as being one of the most promising laboratories in order to understand the origin of holographic duality. In the strict IR limit, the large N version of the theory is believed to be dual to the Lorentzian $ AdS_2 $. As a remarkable feature, the strong coupling ($ |Jt|\gg 1 $) version of the model turns out to be \textit{exactly} solvable at the IR fixed point \cite{Maldacena:2016hyu}-\cite{Rosenhaus:2018dtp}.  However, due to the presence of an infinite tower of \textit{zero} (Goldstone) modes, the corresponding two point correlations diverge and therefore one needs to move slightly away from the deep infrared in order for the theory to make sense. This in turn is related to the fact that quantum gravity in pure $ AdS_2 $ is pathological in the sense that the gravitational back-reaction is infinite. One way to get rid of this instability is to couple the theory with dilaton that eventually breaks some of the IR conformal symmetries of $ AdS_2 $.

The quest for a sensible gravity dual \cite{Teitelboim:1983ux}-\cite{jensen} for the SYK model has been a non trivial issue until very recently \cite{Das:2017pif}-\cite{Lala:2018yib}. The zero temperature version of the model turns out to be dual to a 3D spacetime ($ AdS_2 \times S^1/Z_2 $) that could be viewed as near horizon limit of a near extremal black brane\footnote{For generic $ q $ point vertices this interpretation is not quite obvious \cite{Das:2017hrt}. }. In this model, the dilaton plays the role of the compact third direction ($ S^1/Z_2 $) in the dual gravity picture. It turns out that the spectrum associated with the Kaluza-Klein (KK) modes of the scalar fluctuations (propagating in the bulk 3D spacetime) precisely matches to that with corresponding spectrum of bi-local propagators in the $ q $ SYK model at strong coupling \cite{Das:2017hrt}. The purpose of the present article is to understand as well as extend the notion of $ q $ SYK/AdS duality \cite{Das:2017hrt} in a broader context namely in the presence of Yang-Baxter (YB) deformations \cite{Klimcik:2002zj}-\cite{Klimcik:2008eq}.

The motivation behind the present analysis stems from a recent observation where based on the notion of the YB deformations, the authors in \cite{Kyono:2017jtc}-\cite{Okumura:2018xbh} had proposed an interesting modification of the $ 2D $ dilaton gravity model \cite{Teitelboim:1983ux}-\cite{Almheiri:2014cka} where the quadratic dilaton potential is replaced by a hyperbolic function. The resulting (deformed $ (AdS_2)_{\eta} $) metric turns out to be a solution for the (YB) deformed version of the Almheiri-Polchinski (AP) model \cite{Almheiri:2014cka}. Based on the notion of the holographic correspondence one might therefore tempted to claim that these YB form of deformations \cite{Kyono:2017jtc}-\cite{Okumura:2018xbh} must have a natural interpretation in terms of bi-local excitation \cite{Jevicki:2016bwu}-\cite{Jevicki:2016ito} associated with the large N version of the $ q $ SYK model at strong coupling. With this vision in mind, the goal of the present calculation is therefore to take \textit{holographic} lessons on the implications of YB deformations towards the strong coupling dynamics associated with $ q $ SYK model\footnote{For the sake of technical simplicity, in the present analysis we focus only on \textit{bound} states \cite{Polchinski:2016xgd} with discrete energy eigen values.}. 

The organization of the rest of the paper is the following. In Section 2, we propose a dual gravity picture for the $ q $ SYK model in the presence of YB deformations. In Section 3, we compute two point correlations in the large $ q(\gg 1) $ limit. The strong coupling propagator corresponding to zero modes receives an expected divergent $ \mathcal{O}(J) $ contribution as in the undeformed scenario \cite{Das:2017pif}-\cite{Das:2017hrt}. However, the subleading corrections due to YB effects turn out to be suppressed compared to that with the leading (large $ J $) contribution. Finally, we conclude in Section 4. 
\section{The deformed background}
\subsection{The $ q=4 $ case}
Let us first focus on the background dual to $ q $ SYK model with $ q=4 $ \cite{Das:2017pif},
\begin{eqnarray}
ds^2 = \frac{1}{z^2}\left(-dt^2 +dz^2 \right)+\left(1+\frac{\alpha}{z} \right)^{2}dy^2 
\label{ee2.1}
\end{eqnarray}
which by means of the following substitution,
\begin{eqnarray}
y=\sin^{-1}(\sqrt{\theta})\label{e2.2}
\end{eqnarray}
reverts back the metric of the following form\footnote{Here, the compact direction is constrained by $ 0<| \theta | <1 $. Also notice that unlike the previous analysis \cite{Das:2017hrt} here one needs to retain terms associated with the SYK coupling $ \alpha (\sim 1/J) $ as YB effects manifest at quadratic order in the coupling \cite{Lala:2018yib}. } \cite{Das:2017hrt},
\begin{eqnarray}
ds^2 = \frac{1}{z^2}\left(-dt^2 +dz^2 \right)+\frac{\left( 1+\frac{\alpha}{z}\right)^{2}}{4|\theta | (1-| \theta |)} d \theta^2.
\end{eqnarray}

The above solution (\ref{ee2.1}) could be easily generalized in the presence of YB deformations\footnote{It has been recently argued that (\ref{ee2.4}) could be systematically uplifted to a $ 3D $ geometry \cite{Lala:2018yib} which acts as a solution for the (YB) deformed AP model \cite{Kyono:2017jtc} that replaces the standard dilaton potential with a hyperbolic function. },
\begin{eqnarray}
ds^2_{\eta}=\frac{\mathsf{F}(z , \alpha)}{z^2}(-dt^2 + dz^2)+\left( 1+\frac{1}{2 \eta}\log\left( \frac{1+\eta \alpha / z}{1-\eta \alpha /z}\right) \right)^{2}dy^2 \label{ee2.4}
\end{eqnarray}
where we have introduced the function\footnote{Notice that here we have considered the special case with homogeneous CYBE which amounts of seting, $ X.P =\frac{\alpha}{z} $, $ X^2 =-1 $ and $ P^2=0 $ \cite{Kyono:2017jtc}.},
\begin{eqnarray}
 \mathsf{F}(z , \alpha) = \frac{1}{1-\frac{\eta^2 \alpha^2}{z^2}} .
\end{eqnarray} 

Using (\ref{e2.2}), this trivially generalizes to,
\begin{eqnarray}
ds^2_{\eta}=\frac{\mathsf{F}(z, \alpha)}{z^2}(-dt^2 + dz^2)+\frac{\mathsf{G}^{2}(z ,\alpha)}{4|\theta | (1-| \theta |)} d \theta^2.
\end{eqnarray}
where we have used a short hand notation for the function, 
\begin{eqnarray}
\mathsf{G}(z , \alpha)=1+\frac{1}{2 \eta}\log\left( \frac{1+\eta \alpha / z}{1-\eta \alpha /z}\right).
\end{eqnarray}  
\subsection{The generic $ q$ solution}
We now intend to construct the deformed AP model corresponding to \textit{generic} $ q $ point vertex. To start with, we note down the $ q $ fermion generalization of (\ref{e2.1}),
\begin{eqnarray}
ds^2 =(\sin y)^{2(4/q -1)} \left[ \frac{1}{z^2}\left(-dt^2 +dz^2 \right)+\left(1+\frac{\alpha}{z} \right)^{2}dy^2 \right].
\label{e2.1}
\end{eqnarray}

Using (\ref{e2.2}), this precisely reduces to \cite{Das:2017hrt},
\begin{eqnarray}
ds^2 = | \theta |^{4/q -1}\left[ \frac{1}{z^2}\left(-dt^2 +dz^2 \right)+\frac{\left( 1+\frac{\alpha}{z}\right)^{2}}{4|\theta | (1-| \theta |)} d \theta^2 \right] .\label{e2.7}
\end{eqnarray}

Like in the previous example, a straightforward YB generalization of (\ref{e2.7}) is indeed quite trivial,
\begin{eqnarray}
ds^2_{\eta}= | \theta |^{4/q -1}\left[ \frac{\mathsf{F}(z , \alpha)}{z^2}(-dt^2 + dz^2)+\frac{\mathsf{G}^{2}(z ,\alpha)}{4|\theta | (1-| \theta |)} d \theta^2\right] \label{e2.8}
\end{eqnarray}
which serves as the starting point for our subsequent analysis.
\section{Correlators in the deformed AP model}
\subsection{The quadratic action}
We start considering a single scalar field over the background (\ref{e2.8}),
\begin{eqnarray}
S_{\varphi}=\frac{1}{2}\int dt dz d \theta \sqrt{-g}\mathcal{L}_{\varphi}
\end{eqnarray}
where, we define the corresponding scalar Lagrangian as,
\begin{eqnarray}
\mathcal{L}_{\varphi} = -(\partial_a \varphi)^2 -\mathsf{V}(\theta)\varphi^{2}
\end{eqnarray}
with the scalar potential \cite{Das:2017hrt} introduced as\footnote{Here, $ V $ is some unknown parameter of the model that would be eventually fixed under certain specific physical considerations. },
\begin{eqnarray}
\mathsf{V}(\theta)= \frac{1}{| \theta |^{4/q -1}}\left[ 4 \left(\frac{1}{q}-\frac{1}{4} \right)^{2}+m_0^{2}+\frac{2V}{\mathsf{J}(\theta)}\left(1-\frac{2}{q} \right)\delta(\theta)  \right].
\end{eqnarray}
Notice that here $ V $ is a constant together with,
\begin{eqnarray}
\mathsf{J}(\theta)=\frac{| \theta |^{2/q -1}}{2\sqrt{1-| \theta |}}.
\end{eqnarray}

A straightforward calculation reveals,
\begin{eqnarray}
\sqrt{-g}\mathcal{L}_{\varphi} = \mathsf{J}(\theta)\mathsf{G}(z , \alpha)\left[ (\partial_t \varphi)^{2}-(\partial_z \varphi)^{2}-\frac{4\mathsf{F}(z, \alpha)}{z^2\mathsf{G}^{2}(z , \alpha)}| \theta | (1- | \theta |)(\partial_\theta \varphi)^{2}\right] \nonumber\\
-\frac{\mathsf{J}(\theta)}{z^2}\mathsf{F}(z, \alpha)\mathsf{G}(z , \alpha)\left[ 4 \left(\frac{1}{q}-\frac{1}{4} \right)^{2}+m_0^{2}+\frac{2V}{\mathsf{J}(\theta)}\left(1-\frac{2}{q} \right)\delta(\theta)  \right]\varphi^{2}.
\end{eqnarray}
The range for $ \theta $ integral ranges between -1 to +1 which clearly hits the delta function discontinuity near $ \theta \sim \varepsilon \sim 0 $. One way to get rid of this delta function discontinuity is to impose constraints on the scalar field near this singularity \cite{Das:2017hrt} which is subjected to the Dirichlet boundary conditions of the following form,
\begin{eqnarray}
\varphi (t ,z, \pm 1)=0.\label{e2.16}
\end{eqnarray} 

A careful analysis on the $ \theta $ integral yields the following,
\begin{eqnarray}
S_{\varphi}=S_{\varphi}^{(B)}-\int_{\partial\Sigma} dt dz \frac{\mathsf{F}(z, \alpha)}{z^2 \mathsf{G}(z , \alpha)}\varphi (t,z , 0)\mathcal{B}(t,z,\varepsilon)
\end{eqnarray}
where, we set the constraint,
\begin{eqnarray}
 \mathcal{B}(t,z,\varepsilon) = \left[2\theta^{2/q}\partial_{\theta}\varphi \right]_{\theta =\varepsilon}+ V\mathsf{G}^{2}(z, \alpha)\varphi (t,z,0)=0\label{e2.18}
\end{eqnarray}
subjected to the fact that, $ \varphi (t,z, \varepsilon)= \varphi (t,z, -\varepsilon)$ \cite{Das:2017hrt}. This finally leads to the bulk quadratic action of the following form,
\begin{eqnarray}
S^{(B)}_{\varphi}=\frac{1}{2}\int d^2x \int_{0}^{1}d \theta \mathsf{J}(\theta)\varphi (x ,\theta) \hat{\mathcal{D}}\varphi (x, \theta);~(x=t,z)
\end{eqnarray}
where, we have introduced the operator,
\begin{eqnarray}
 \hat{\mathcal{D}}=\mathsf{G}(z,\alpha)(-\partial^2_{t}+\partial^{2}_z)+\mathsf{G}'(z,\alpha)\partial_{z}+\frac{4 \mathsf{F}(z,\alpha)}{z^2 \mathsf{G}(z,\alpha)}\left[\theta (1-\theta)\partial^2_{\theta} +\left(\frac{2}{q}-\theta \left(\frac{1}{2}+\frac{2}{q} \right)  \right)\partial_{\theta} \right] \nonumber\\
 -\frac{\mathsf{F}(z, \alpha)}{z^2}\mathsf{G}(z, \alpha)\left[ 4 \left(\frac{1}{q}-\frac{1}{4} \right)^{2}+m_0^{2} \right].\nonumber\\
 \label{e2.20}
\end{eqnarray}

Considering a strong coupling limit ($ |J| \gg 1 $) it is quite natural to expand (\ref{e2.20}) perturbatively in the coupling $ \alpha (\sim 1/J) $ \cite{Das:2017pif} which yields the following expansion of the operator,
\begin{eqnarray}
 \hat{\mathcal{D}}= \hat{\mathcal{D}}^{(0)}+ \delta\hat{\mathcal{D}}+\mathcal{O}(\alpha^3)\label{e2.21}
\end{eqnarray}
where, the individual terms in the expansion (\ref{e2.21}) could be formally expressed as,
\begin{eqnarray}
 \hat{\mathcal{D}}^{(0)}=-\partial^2_{t}+\partial^{2}_z+\frac{4}{z^2}\left[\theta (1-\theta)\partial^2_{\theta} +\left(\frac{2}{q}-\theta \left(\frac{1}{2}+\frac{2}{q} \right)  \right)\partial_{\theta} \right]\nonumber\\
 -\frac{1}{z^2}\left[ 4 \left(\frac{1}{q}-\frac{1}{4} \right)^{2}+m_0^{2} \right]\label{e2.22}
\end{eqnarray}
\begin{eqnarray}
 \delta\hat{{\mathcal{D}}}=\frac{\alpha}{z}\left[ (-\partial^2_{t}+\partial^{2}_z)-\frac{1}{z}\partial_z \right]-\frac{\alpha }{z^3}\left(1+\frac{  \eta ^2 \alpha}{z} \right) \left[ 4 \left(\frac{1}{q}-\frac{1}{4} \right)^{2}+m_0^{2} \right]\nonumber\\ -\frac{4 \alpha }{z^3}\left(1-\frac{ \alpha  \left(\eta ^2+1\right)}{z} \right) \left[\theta (1-\theta)\partial^2_{\theta} +\left(\frac{2}{q}-\theta \left(\frac{1}{2}+\frac{2}{q} \right)  \right)\partial_{\theta} \right].
 \label{e2.23}
\end{eqnarray}

Notice that (\ref{e2.22}) is identical to that with the earlier results in \cite{Das:2017hrt}. However on the other hand, we identify (\ref{e2.23}) as the new addition to the SYK spectrum that has its root in YB deformations associated with the $ AdS_2 $ sigma model\footnote{At the level of the $ AdS_2 $ sigma model, the YB deformation is introduced in terms of a classical $ \mathfrak{R} $ operator $ S_{YB}\sim \int d^2 \sigma \gamma^{\alpha \beta} Tr (J_{\alpha}\frac{1}{(1 - 2\eta \mathfrak{R} \circ P)}P(J_{\beta}) )$ where $ P $ is the projection. Here, $ J_{\alpha}=g^{-1}\partial_{\alpha}g $ is the usual left invariant one form constructed out of the $ AdS_2 $ coset representative $ g\in SL(2)/U(1) $. For the purpose of our present analysis we restrict ourselves to the special choice of classical $ \mathfrak{R} $ operator that satisfies \textit{homogeneous} classical YB equation.  } \cite{Kyono:2017jtc}-\cite{Okumura:2018xbh}. Going into the Fourier space, one might express the scalar modes as\footnote{In principle the integral over $ \nu $ could be thought of as being that of the sum over both discrete (bound states) modes as well as the continuous modes/scattering states \cite{Polchinski:2016xgd}. However, for the purpose of our present analysis we would like to restrict our computations only for those of the discrete modes with $ \nu=3/2+2n~ (n=0,1,2..) $ that also sets, $ \mathcal{N}_{\nu}=(2 \nu)^{-1} $ \cite{Jevicki:2016bwu}. Therefore one might thought of replacing the integral as a sum over discrete modes and/ or the bound states, $$ \int d\nu \rightarrow \sum_{\nu=3/2 +2n} $$.},
\begin{eqnarray}
\varphi (x ,\theta)&=&\int \frac{dw d\nu dk}{\mathcal{N}_{\nu}}e^{-i w t}\sqrt{z}\mathcal{Z}_{\nu}(|wz|)\sigma_{k}(\theta)\zeta_{w} ( \nu, k)
\end{eqnarray}
which finally yields the quadratic action of the following form,
\begin{eqnarray}
S^{(B)}_{\varphi}=\frac{1}{2}\int dt dz d\theta \mathsf{J}(\theta)\int \frac{dw'  d\nu' dk' }{\mathcal{N}_{\nu} \mathcal{N}_{\nu'}}e^{-iw' t}\sqrt{z}\mathcal{Z}^{\ast}_{\nu'}(| w' z |)\sigma_{k'}(\theta)\zeta_{w'} ( \nu', k')\nonumber\\
\times \int dw d\nu dk\hat{\mathcal{D}}e^{-iwt}\sqrt{z}\mathcal{Z}_{\nu}(| wz |)\sigma_{k}(\theta) \zeta_{w} ( \nu, k).\label{E2.25}
\end{eqnarray}

Our next task would be to diagonalize the operator (\ref{e2.21}), $ \hat{\mathcal{D}}(=  \hat{\mathcal{D}}^{(0)} +  \hat{\mathcal{D}}^{(1)} )$ and obtain the corresponding spectra. This eventually translates into a problem of solving the corresponding eigen value equation associated with this operator. Before we proceed further, it is customary to express the operator (\ref{e2.21}) in the following form,
\begin{eqnarray}
\hat{\mathcal{D}} = \hat{\mathcal{D}}_{z}(\alpha , \eta , z)+\frac{4}{z^2} \mathcal{F}(\alpha , \eta,  z) \hat{\mathcal{D}}_{\theta}(\theta) \label{e2.24}
\end{eqnarray}
where, the individual components could be formally expressed as,
\begin{eqnarray}
\hat{\mathcal{D}}_{z} = \left( 1+\frac{\alpha}{z}\right)(\partial^{2}_z-\partial_{t}^2) -\frac{\alpha}{z^2}\partial_{z}
-\frac{\mu^{2}}{z^2}
\end{eqnarray}
\begin{eqnarray}
\hat{\mathcal{D}}_{\theta}(\theta)= \theta (1-\theta)\partial^2_{\theta} +\left(\frac{2}{q}-\theta \left(\frac{1}{2}+\frac{2}{q} \right)  \right)\partial_{\theta}
\end{eqnarray}
together with the function,
\begin{eqnarray}
\mathcal{F}&=& 1-\frac{\alpha}{z}\left(1-\frac{ \alpha  \left(\eta ^2+1\right)}{z} \right)\nonumber\\
\mu^2 &=&\mu_0^2 (q) \left( 1+\frac{\alpha}{z}\left(1+\frac{  \eta ^2 \alpha}{z} \right)\right);~\mu^2_0 (q) =m^2_{0}+4\left( \frac{1}{q}-\frac{1}{4}\right)^{2}.
\end{eqnarray}
\subsection{Diagonalization of $ \hat{\mathcal{D}}_{\theta} $}
We start with proposing the eigenvalue equation corresponding to the angular operator,
\begin{eqnarray}
 \hat{\mathcal{D}}_{\theta}(\theta)\sigma_{k}(\theta) = -\frac{k^{2}}{4}\sigma_{k} (\theta)\label{e2.32}
\end{eqnarray}
which has a solution of the following form,
\begin{eqnarray}
\sigma_{k}(\theta)=\mathcal{C}_{1}  (-1)^{\frac{q-2}{q}} \theta ^{\frac{q-2}{q}} \, _2F_1\left(a_1,a_2;2-\frac{2}{q};\theta \right)
+\mathcal{C}_{2} \, _2F_1\left(b_1,b_2;\frac{2}{q};\theta \right)\label{e2.33}
\end{eqnarray}
where, the individual entities could be formally expressed as,
\begin{eqnarray}
a_{n}=\frac{(-1)^{n}}{4q}(\sqrt{q \left(4 q k^2+q-8\right)+16}-(-1)^{n}(4-3q)),\nonumber\\
b_n=-\frac{(-1)^{n}}{4q}(\sqrt{q \left(4 k^2 q+q-8\right)+16}+(-1)^n(q-4)).
\end{eqnarray}
Here, $ \mathcal{C}_{1,2} $ are integration constants that needs to be fixed using boundary conditions. 

Substituting the above solution (\ref{e2.33}) into (\ref{e2.18}) and subsequently expanding in the small variable $ \theta (\sim 0) $ one finds\footnote{In order to arrive at the above condition (\ref{e2.35}) we have made use of the fact that the parameter $ q(\gg 1) $ (together with the fact $ | q\theta |<1 $) of the theory is large enough so that one can consider an expansion in $ 1/q $ and keep terms only upto leading order in the expansion.}, 
\begin{eqnarray}
\mathcal{C}_1 +\frac{k^2 q}{8}\mathcal{C}_2 \approx \frac{V}{2}\mathcal{C}_2. \label{e2.35}
\end{eqnarray}

On the other hand, imposing the other boundary condition (\ref{e2.16}) one finds,
\begin{eqnarray}
\mathcal{C}_{1}\approx \frac{\mathcal{C}_{2}}{32}(q+\pi +6 \log 2).\label{e2.36}
\end{eqnarray}

Combibing (\ref{e2.35}) and (\ref{e2.36}) we find,
\begin{eqnarray}
V= \frac{k^2 q}{4}+\frac{1}{16}(q+\pi +6 \log 2)+\mathcal{O}(1/q).
\end{eqnarray}

Setting\footnote{This solution is obtained (in the large $ q(\gg 1) $ limit) by demanding that the leading order propagator (\ref{E4.16}) associated to zero modes has a pole.},
\begin{eqnarray}
k^2 \simeq 2-\frac{4}{q^2}
\end{eqnarray}
for \textit{zero} modes in the large $ q $ limit we finally fix the potential,
\begin{eqnarray}
V(q)= \frac{9 q}{16}+\pi +6 \log 2+\mathcal{O}(1/q).
\end{eqnarray}
\subsection{ The Green's function}
The action (\ref{E2.25}) could be expressed as a perturbation in the coupling $ \alpha $ namely,
\begin{eqnarray}
S^{(B)}_{\varphi}=S^{(0)}_{\varphi}+\alpha S^{(1)}_{\varphi}+\alpha^2 S^{(2)}_{\varphi}.\label{e4.11}
\end{eqnarray}
In the following, we evaluate each of these above entities separately. 
\subsubsection{Zeroth order computation}
We first note down the action at zeroth order in $ \alpha $,
\begin{eqnarray}
S^{(0)}_{\varphi}=\frac{1}{2}\int dt dz d\theta \mathsf{J}(\theta)\int \frac{dw'  d\nu' dk' }{\mathcal{N}_{\nu} \mathcal{N}_{\nu'}}e^{-iw' t}\sqrt{z}\mathcal{Z}^{\ast}_{\nu'}(| w' z |)\sigma_{k'}(\theta)\zeta_{w'} ( \nu', k')\nonumber\\
\times \int dw d\nu dk\hat{\mathcal{D}}^{(0)}e^{-iwt}\sqrt{z}\mathcal{Z}_{\nu}(| wz |)\sigma_{k}(\theta) \zeta_{w} ( \nu, k)
\end{eqnarray}
where, the zeroth order operator could be expressed as usual,
\begin{eqnarray}
\hat{\mathcal{D}}^{(0)}=-\partial^{2}_{t}+\partial^{2}_{z}-\frac{\mu^2_{0}(q)}{z^2}+\frac{4}{z^2}\hat{\mathcal{D}}_{\theta}(\theta) .
\end{eqnarray}

Using the completeness relation for the Bessel  function together with the orthogonality condition \cite{Das:2017hrt} for $\sigma_{k}(\theta)  $ it is indeed trivial to show,
\begin{eqnarray}
S^{(0)}_{\varphi}=\frac{1}{2}\int dw d\nu dk \frac{\mathcal{C}(k)}{\mathcal{N}_{\nu}}\zeta_{-w}(\nu ,k)(\nu^2 -\nu^{2}_{0}(q))\zeta_{w}(\nu ,k)\label{E4.16}
\end{eqnarray}
where, we have introduced,
\begin{eqnarray}
\nu^{2}_{0}(q)=\mu^{2}_{0}(q)+k^2 +\frac{1}{4}.
\end{eqnarray}

It is noteworthy to mention that $ \mathcal{Z}_{\nu}(| wz |) $ satisfies the standard Bessel equation,
\begin{eqnarray}
(z^2 \partial^{2}_{z}+z\partial_{z}+w^2 z^2)\mathcal{Z}_{\nu}(| wz |)=\nu^2 \mathcal{Z}_{\nu}(| wz |).\label{e4.16}
\end{eqnarray}
\subsubsection{Yang-Baxter shift}
The first order shift in the action (\ref{e4.11}) could be formally expressed as,
\begin{eqnarray}
S^{(1)}_{\varphi}=\frac{1}{2}\int dt dz d\theta \mathsf{J}(\theta)\int \frac{dw'  d\nu' dk' }{\mathcal{N}_{\nu} \mathcal{N}_{\nu'}}e^{-iw' t}\sqrt{z}\mathcal{Z}^{\ast}_{\nu'}(| w' z |)\sigma_{k'}(\theta)\zeta_{w'} ( \nu', k')\nonumber\\
\times \int dw d\nu dk\hat{\mathcal{D}}^{(1)}e^{-iwt}\sqrt{z}\mathcal{Z}_{\nu}(| wz |)\sigma_{k}(\theta) \zeta_{w} ( \nu, k)\label{e4.17}
\end{eqnarray}
where, the operator above in (\ref{e4.17}) is given by,
\begin{eqnarray}
\hat{\mathcal{D}}^{(1)}=\frac{1}{z}(-\partial^{2}_{t}+\partial^{2}_{z})-\frac{1}{z^2}\partial_{z}-\frac{\mu^2_{0}(q)}{z^3}-\frac{4}{z^3}\hat{\mathcal{D}}_{\theta}(\theta).
\end{eqnarray}

Using (\ref{e4.16}), it is quite straightforward to show,
\begin{eqnarray}
\hat{\mathcal{D}}^{(1)}e^{-iwt}\sqrt{z}\mathcal{Z}_{\nu}(| wz |)\sigma_{k}(\theta)=\frac{1}{z^{5/2}}\left(-z\partial_{z}+\mathcal{G}(\nu ,q,k) \right) e^{-iwt}\mathcal{Z}_{\nu}(| wz |)\sigma_{k}(\theta)\label{e4.19}
\end{eqnarray}
where, the new entity above could be expressed as,
\begin{eqnarray}
\mathcal{G}(\nu ,q, k)= \nu^2 -\frac{3}{4}-\mu^2_{0}(q)+k^2.
\end{eqnarray}

Using the following identity \cite{Das:2017pif},
\begin{eqnarray}
\partial_{z}\mathcal{Z}_{\nu}(|wz|)=\frac{\nu}{| z |}\mathcal{Z}_{\nu}(|wz|) -| w | (J_{\nu +1}(|w z|)-\xi_{\nu}J_{-\nu -1}(|w z|))
\end{eqnarray}
 one might further simplify (\ref{e4.19}) as\footnote{Here, we have summed over only real modes with discrete energy eigen values with, $ \nu =3/2+2n $\cite{Polchinski:2016xgd}.},
 \begin{eqnarray}
 \hat{\mathcal{D}}^{(1)}e^{-iwt}\sqrt{z}\mathcal{Z}_{\nu}(| wz |)\sigma_{k}(\theta)=\left[ \frac{1}{z^{5/2}}(\mathcal{G}(\nu ,q, k)-\nu)\mathcal{Z}_{\nu}(|wz|)+\frac{|w|}{z^{3/2}}\mathcal{Z}_{\nu +1}(|wz|)\right] e^{-iwt}\sigma_{k}(\theta).\label{e4.22}\nonumber\\
 \end{eqnarray}

Substituting (\ref{e4.22}) into (\ref{e4.17}) we find,
\begin{eqnarray}
S^{(1)}_{\varphi}=\Delta_{1}+\Delta_{2}
\end{eqnarray}
where, the individual entities could be formally expressed as,
\begin{eqnarray}
\Delta_{1}&=&-\frac{1}{2}\int dw dk d\nu d\nu' \frac{\mathcal{C}(k)|w|}{\mathcal{N}_{\nu}\mathcal{N}_{\nu'}}\zeta_{-w}(\nu' ,k)\mathcal{H}_{1}(\nu ,\nu' , q)\zeta_{w}(\nu ,k)\nonumber\\
\mathcal{H}_{1}&=&\frac{4 \cos \left(\frac{1}{2} \pi  (\nu' -\nu )\right)(\mathcal{G}(\nu ,q, k)-\nu)}{\pi  \left(\nu^{'4}-2 \nu^{'2} \left(\nu ^2+1\right)+\left(\nu ^2-1\right)^2\right)}
\end{eqnarray}
and,
\begin{eqnarray}
\Delta_{2}&=&\frac{1}{2}\int dw dk d\nu d\nu' \frac{\mathcal{C}(k)|w|}{\mathcal{N}_{\nu}\mathcal{N}_{\nu'}}\zeta_{-w}(\nu' ,k)\mathcal{H}_{2}(\nu ,\nu' )\zeta_{w}(\nu ,k)\nonumber\\
\mathcal{H}_{2}&=&\frac{2 \cos \left(\frac{1}{2} \pi  (\nu' -\nu )\right)}{\pi  \left((\nu +1)^2-\nu ^{'2}\right)}.
\end{eqnarray}

Finally, we note down the correction at quadratic order in the coupling,
\begin{eqnarray}
S^{(2)}_{\varphi}&=&\frac{1}{2}\int dw dk d\nu d\nu' \frac{\mathcal{C}(k)|w|}{\mathcal{N}_{\nu}\mathcal{N}_{\nu'}}\zeta_{-w}(\nu' ,k)\Sigma(\nu ,\nu' , q)\zeta_{w}(\nu ,k)\nonumber\\
\Sigma &=&\frac{16 \sin \left(\frac{1}{2} \pi  (\nu' -\nu )\right)((\mu^2_{0}(q)+k^2)\eta^2 +k^2)}{\pi  ( (\nu' -\nu)^2 -4)(\nu^{'2} -\nu^2 )  ( (\nu' +\nu)^2 -4)}.
\end{eqnarray}

We focus on the zero modes \cite{Das:2017pif} with $ \nu =\nu' =\frac{3}{2} $ and set the scalar mass at the BF bound, $ m^2_{0}=-\frac{1}{4} $ which finally yields the two point correlation in the momentum space,
\begin{eqnarray}
\langle \zeta_{-w}(3/2 ,k\sim \sqrt{2}) \zeta_{w}(3/2 ,k \sim \sqrt{2})\rangle_{q\gg 1} \approx -\frac{9 \pi}{4 \mathcal{C}(\sqrt{2}) |w|\Gamma (\alpha)}\label{e4.29}
\end{eqnarray}
where we have defined,
\begin{eqnarray}
\Gamma (\alpha) = \alpha +\frac{8 \pi}{15}  \alpha^{2} (\eta^2 +1)+\mathcal{O}(\alpha^3).\label{e4.30}
\end{eqnarray}
The second term on the R.H.S. of (\ref{e4.30}) is precisely the YB contribution to the spectrum of zero modes associated with the $ q $ SYK model in the limit of large $ q $. 

Using (\ref{e4.29}) we finally express the space time propagator as,
\begin{eqnarray}
\langle \Phi (x ,\theta)\Phi (x', \theta')\rangle_{q \gg 1}& \approx &-\frac{9\pi}{4\Gamma (\alpha)}| z z' |^{1/2}\Theta (\theta ,\theta')\int \frac{dw}{| w |}e^{-i w (t-t')}\mathcal{Z}_{3/2}(|w z|)\mathcal{Z}_{3/2}(|w z'|)\nonumber\\
\Theta (\theta ,\theta') & =&\frac{\sigma_{k}(\theta)\sigma_{k}(\theta')}{ \mathcal{C}(k)}\Big|_{k^2 \sim 2+\mathcal{O}(1/q^2)}.
\end{eqnarray}
A straightforward computation reveals,
\begin{eqnarray}
\sigma_k (0)^2\Big|_{k^2 \sim 2+\mathcal{O}(1/q^2)} \simeq \mathcal{C}^{2}_{1}
\end{eqnarray}
which thereby yields the space time propagator,
\begin{eqnarray}
\langle \Phi (x ,0)\Phi (x', 0)\rangle_{q \gg 1} \approx -\frac{9 \pi \mathcal{C}^{2}_{1}}{4 \alpha ~\mathcal{C}(k)}\left( 1-\frac{8 \pi}{15}  \alpha (\eta^2 +1)\right) | z z' |^{1/2}~~~\nonumber\\
\times \int \frac{dw}{ | w |}e^{-i w (t-t')}\mathcal{Z}_{3/2}(|w z|)\mathcal{Z}_{3/2}(|w z'|)\label{e4.33}
\end{eqnarray}
 together with,
 \begin{eqnarray}
 \mathcal{C}(k)=\int_{0}^{1}d\theta \mathsf{J}(\theta)\sigma^{2}_{k}(\theta).
 \end{eqnarray}
 
 Notice that (\ref{e4.33}) is precisely of the form \cite{Maldacena:2016hyu} and is sort of expected from the earlier analysis \cite{Das:2017pif}. The leading order contribution goes as $ \mathcal{O}(J) $ \cite{Das:2017pif} in the limit of large $ J $ which thereby diverges. However, the sub-leading (YB) contributions turn out to be suppressed (compared with the leading term) in the limit of strong coupling. 
\section{Summary and final remarks}
Before we conclude this Letter, several important remarks are in order. In the very first place, one should notice that the above background (\ref{e2.8}) suffers from an unusual metric singularity associated with the function $ \mathsf{F}(z , \alpha) $. This constraints our bulk calculations within a finite radial cutoff,
\begin{eqnarray}
|z_B| =\eta~\alpha \equiv \frac{\eta}{J}.\label{eee3.11}
\end{eqnarray}

The existence of such radial cutoffs turns out to be a very special feature for spacetimes embodied with YB deformations \cite{Kameyama:2014vma}-\cite{yoshida}. One could therefore imagine constraining the bulk spacetime within a singularity surface (also known as the \textit{holographic screen} \cite{Kameyama:2014vma}-\cite{yoshida}) whose radial location is fixed at a finite distance $ z=z_B $ that eventually could be treated as the \textit{boundary} for the bulk spacetime under consideration. The dual QFT is therefore considered to be living on this holographic screen. 

As far as the Gauge/Gravity duality is concerned, the above entity (\ref{eee3.11}) could be thought of as being that of the energy scale associated with the holographic RG flow. In other words, depending on the (radial) location of the singularity surface one is supposed to probe the dynamics associated with the dual QFT under the RG flow. 
The UV fixed point associated with this RG flow corresponds to the criteria, $| \frac{\eta}{J}| \sim 0$ which yields the metric corresponding to $ AdS_2 $ with dilaton being completely trivial. On the other hand, the end point of this RG flow is determined by the criteria, $ |\frac{\eta}{J}|=\Lambda_{IR}\gg 1 $ where $ \Lambda_{IR} $ stands for that of the deep IR cutoff in the bulk. 
 Therefore in the limit, $ | \frac{\eta}{J}|\ll 1 $ (which is the limit considered in this paper) quantum $ \mathcal{O} $ ($1/\sqrt{N}$) modifications due to YB deformations should be regarded as being that of the reminiscent of some high energy (UV) physics over the critical (saddle point) IR background \cite{Lala:2018yib}. However, as expected, in the strict limit of $ \frac{\eta}{J}=0 $ these UV effects completely decouple from the rest of the spectrum and one recovers the original ``IR physics" corresponding to the $ q $ SYK model.
\section*{Acknowledgments}
The author is indebted to the authorities of IIT Roorkee for their unconditional support towards researches in
basic sciences.

\end{document}